\documentclass[aps,prl,amssymb,twocolumn,superscriptaddress,showpacs]{revtex4}

\usepackage{graphicx}
\usepackage{dcolumn}
\usepackage{bm}

\begin{document}

\newcommand{\beq}{\begin{equation}}
\newcommand{\eeq}{\end{equation}}
\newcommand{\barr}{\begin{eqnarray}}
\newcommand{\earr}{\end{eqnarray}}

\newcommand{\andy}[1]{ }

\newcommand{\bmsub}[1]{\mbox{\boldmath\scriptsize $#1$}}

\def\bra#1{\langle #1 |}
\def\ket#1{| #1 \rangle}
\def\sinc{\mathop{\text{sinc}}\nolimits}
\def\cV{\mathcal{V}}
\def\cH{\mathcal{H}}
\def\cT{\mathcal{T}}
\renewcommand{\Re}{\mathop{\text{Re}}\nolimits}

\title{Interference of mesoscopic particles: quantum--classical transition}

\author{P. Facchi}
\affiliation{Dipartimento di Matematica, Universit\`a di Bari,
        I-70125  Bari, Italy}
\affiliation{INFN, Sezione di Bari, I-70126 Bari, Italy}
\author{S. Pascazio} \affiliation{Dipartimento di Fisica,
Universit\`a di Bari,
        I-70126  Bari, Italy}
\affiliation{INFN, Sezione di Bari, I-70126 Bari, Italy}
\author{T. Yoneda}
\affiliation{Dipartimento di Fisica, Universit\`a di Bari,
        I-70126  Bari, Italy}
\affiliation{School of Medical Sciences, Kumamoto University,
4-24-1 Kuhonji, 862-0976 Kumamoto, Japan }

\date{\today}

\begin{abstract}
We analyze the double slit interference of a mesoscopic particle.
We calculate the visibility of the interference pattern, introduce
a characteristic temperature that defines the onset to decoherence
and scrutinize the conditions that must be satisfied for an
interference experiment to be possible.
\end{abstract}


\maketitle

\underline{Introduction} - Interference is one of the most
characteristic traits of quantum systems. As Dirac clarified
\cite{Dirac}, this phenomenon is rooted in the superposition
principle, according to which different states of a single quantum
mechanical particle interfere with each other. The simplest case
is that of two states: double slit interference has been observed
with photons, electrons, neutrons, atoms and small molecules
\cite{interferencebooks}, and recently even with large molecular
clusters \cite{AZfull}. Our comprehension of the quantum
mechanical world has been shaped, to a large extent, by the ideas
that motivated these experiments.

Quantum particles interfere, but classical particles do not, and
it is not easy to understand where the borderline has to be
placed. The size of the interfering system plays an important
role, but it is certainly not the only relevant variable: for
example, double slit interference has been observed with
molecules, but not with protons. In this Letter we will analyze
the interference of mesoscopic systems, endowed with an internal
structure which leads to entanglement with their environment
(e.g., via photon emission). An example is a fullerene molecule
flying between a diffraction grating and a detector. Some
classical features are apparent for such mesoscopic systems, yet
their ability to interfere is preserved, at least to some extent,
and can be viewed as a quantum signature. The main objective of
this Letter will be to understand under which conditions a
``large" system interferes and which ones of its dynamical
variables can interfere.

\underline{Double slit interference in a Poissonian environment} -
Let us consider a mesoscopic quantum system (molecule), whose
center of mass is described by a (double) wave packet
$|\psi_{0}\rangle$, emerging from a double slit. The wave packet
travels along direction $+z$; the slits are parallel to $y$ and
are separated by a distance $d$, along direction $x$ ($\hbar=1$):
\begin{eqnarray}
|\psi_0 \rangle &=& \frac{1}{\sqrt{2}}(|\psi_{+} \rangle +
|\psi_{-} \rangle),
\nonumber \\
|\psi_{\pm}\rangle &=& \exp\left(\pm i\frac{d}{2}p_x \right)
|\psi_{\rm slit}\rangle ,
\label{eq:initial}
\end{eqnarray}
where $p_x$ is the $x$ component of the momentum operator and
$|\psi_{\rm slit}\rangle$ the state emerging from one slit. We
assume that $\langle\psi_{-} |\psi_{+}\rangle=0$, so that $|\psi_0
\rangle$ is normalized. During its travel to the screen, the molecule emits
photons and recoils accordingly. The internal state of the
molecule together with the photon field plays therefore the role
of environment: such an environment disturbs the motion of the
center of mass, via scattering processes (typically photon
emissions, yielding momentum kicks). We shall assume that the
internal temperature of the mesoscopic system is much higher than
the temperature of the photon field.

Let the molecule undergo momentum kicks ($\triangle p_i, i \in
\mathbb{Z}$) due to photon emissions. The Hamiltonian describing
the evolution of the $x$ component of the center of mass in the
presence of random kicks $\Delta p_{k}$ at times $t_k$ reads
(henceforth, for simplicity, $p_x=p$)
\begin{eqnarray}
 H_{\xi}(t)&=&\frac{p^2}{2m}-\xi(t) x, \label{eq:Hamiltonian1} \\
\xi(t)&=& \sum_{k}\delta(t-t_{k})\Delta p_{k},
\label{eq:Hamiltonian}
\end{eqnarray}
where $\{t_k\}_{k\in \mathbb{Z}}$ is a shot noise with density
$\Lambda$ and the momentum jumps $\Delta p_k$ are independent
identically distributed random variables with probability density
$W(\Delta p_k)$. The process $\xi(t)$ is the time derivative of a
compound Poisson process \cite{Snyder}. Both $\Lambda$ and $W$ are
functions of the state of the environment (for example its temperature
of $T$).

The time evolution of a wave packet which emerges at time $t_0=0$
from the slits reads
\begin{equation}\label{eq:final}
|\psi(t) \rangle = U_{\xi}(t)|\psi_0 \rangle ,
\end{equation}
$U_{\xi} (t)$ being the unitary evolution engendered by the
Hamiltonian (\ref{eq:Hamiltonian})
\begin{eqnarray}
U_{\xi} (t) &=& \cT \exp\left(-i\int_{0}^t ds \; H_{\xi}(s)\right)
\nonumber\\
& = & e^{-i (t-t_n)p^{2}/2m} \cT \prod_{k=1}^n e^{i x\triangle
p_k}
e^{-i \triangle t_{k-1} p^{2}/2m} \nonumber\\
& = & e^{i x \triangle p^{(n)}} \prod_{k=0}^{n} e^{-i (p-\triangle
p^{(k)})^{2}\triangle t_{k}/2m} ,
\label{eq:Uxi}
\end{eqnarray}
where $\triangle t_k=t_{k+1}-t_{k}$, $\triangle
p^{(k)}=\sum_{j=1}^k \triangle p_j$ is the total effect of $k$
momentum jumps ($\triangle p^{(0)}=0$), the total number of
collisions $n$ is a Poisson random variable with mean $\Lambda t$,
and $\cT$ is the time-ordering operator, forcing earlier times
(lower $k$) at the right. In the third equality we used the
commutation relation
$e^{-i p^2 \triangle t/2m} e^{i x \triangle p}=e^{i x \triangle
p}e^{-i (p-\triangle p)^2 \triangle t/2m}$, in order to move all
kick operators to the far left side.

From (\ref{eq:final}) and (\ref{eq:Uxi}) one gets
\begin{eqnarray}
\psi(x,t) &=& \langle x|\psi(t) \rangle
\nonumber \\
&=& e^{i x \triangle p^{(n)}} \bra{x} e^{-i \sum (p-\triangle
p^{(k)})^{2} \triangle t_{k}/2m }\ket{\psi_0}
\nonumber \\
&=& e^{i x \triangle p^{(n)}}\int \frac{dp}{\sqrt{2\pi}} e^{-i
\phi(p)}\tilde \psi_0(p),
\label{eq:diff-wave0}
\end{eqnarray}
where $\tilde \psi_0(p)=\bra{p}\psi_0\rangle$ and
\begin{eqnarray}
\label{eq:phi(p)}
\phi(p)&=&\sum_{k=0}^{n} \frac{(p-\triangle p^{(k)})^{2}\triangle
t_{k}}{2m}-x p
\end{eqnarray}
is a quadratic polynomial in $p$ with quadratic term $p^2\sum
\triangle t_{k}/2m= p^2 t/2m$. It can be rewritten as
\begin{equation}\label{eq:eq:phi(p)1}
\phi(p)= \phi(\bar p) + \frac{t}{2m} (p-\bar p)^2 ,
\end{equation}
$\bar p$ being the value of the momentum at the extremal
$\phi'(\bar p)=0$, that is
\begin{equation}\label{eq:barp}
\bar p (x,t) = \frac{m x}{t} + \sum_{k=0}^{n} \frac{\triangle
p^{(k)}\triangle t_{k}}{t}= \frac{m x}{t} + \sum_{k=1}^{n} \zeta_k
\triangle p_k  ,
\end{equation}
where $\zeta_k = 1- t_k/t$ characterize the emissions between the
grating and screen. A peculiarity of this analysis is the presence
of the \emph{same} Poisson process on \emph{both} branch waves:
for an external environment, one should have considered
\emph{two independent} Poisson processes, one for each branch wave.

Equation (\ref{eq:diff-wave0}) represents the convolution of the
initial momentum wave packet with a Gaussian
\begin{eqnarray}
\psi(x,t) = e^{i\left[x \triangle p^{(n)}-\phi(\bar p)\right]}\int
\frac{dp}{\sqrt{2\pi}} e^{-i \frac{t}{2 m} (p-\bar p)^2}\tilde
\psi_0(p),
\label{eq:diff-wave1}
\end{eqnarray}
whose spread $(m/t)^{1/2}$ becomes narrower as time $t$ increases.
For $t\to\infty$ (\ref{eq:diff-wave1}) reads
\begin{equation}\label{eq:asympt}
\psi(x,t)\sim e^{i\left[x \triangle p^{(n)}-\phi(\bar
p)\right]}\left(\frac{m}{it}\right)^{\frac{1}{2}}\tilde
\psi_0(\bar p).
\end{equation}
This approximation is valid for $t \gg m |
\tilde\psi_0^{\prime\prime} (\bar p)/\tilde\psi_0 (\bar p)|$ and
implies that $\tilde \psi_0(p)$ can be represented by the constant
value $\tilde \psi_0(\bar p)$ on the screen. In the following we
will always suppose that such condition holds (far field
interference pattern). The interference pattern reads
\begin{eqnarray}
 I(x,t)=\left\langle\left|\psi(x,t)\right|^2\right\rangle
 \sim \left(\frac{m}{t }\right) \left\langle|\tilde{\psi}_{0}(\bar
 p)|^2\right\rangle
\label{eq:diff-wave}
\end{eqnarray}
where $\langle \cdots \rangle$ denotes the average over the
process $\xi(t)$. With the initial state (\ref{eq:initial}),
$\tilde \psi_0(p)= \sqrt{2} \tilde\psi_{\rm slit}(p)\cos(pd/2)$,
with $\tilde\psi_{\rm slit}(p)\equiv \langle p|\psi_{\rm
slit}\rangle$, and the far-field condition is satisfied for $t\gg
m d^2$. Under this condition, the intensity at the screen reads
\begin{eqnarray}
I(x,t )= \left(\frac{m}{t }\right) \left|\tilde{\psi}_{\rm slit}
\left(\frac{mx}{t}\right)\right|^2 \left[ 1+ \left\langle
\cos\left(\bar p d\right) \right\rangle \right],
\label{eq:Intensity}
\end{eqnarray}
where we approximated $|\tilde\psi_{\rm slit}(\bar
p)|^2\simeq|\tilde\psi_{\rm slit}(mx/t)|^2 $, for weak enough kicks.
The corresponding visibility $\cV$ is
\begin{eqnarray}
 I(x,t ) &=& I_0(x,t ) \left[ 1+ \cV \cos\left(\frac{md}{\hbar
 t }x+\phi \right)\right] , \nonumber \\
 \cV &=& |F|, \quad \phi = \arg F,
 \nonumber \\
 F&=&\left\langle \exp\left(i d \sum_{k=1}^{n} \zeta_k
\triangle p_k\right)\right\rangle = \cV e^{i\phi},
\label{eq:Vis-Int}
\end{eqnarray}
where $I_0(x,t)=(m/t)|\tilde\psi_{\rm slit}(mx/t)|^2$. In order to
calculate the visibility, the features of the average $\langle
\cdots \rangle$ must be expressed in terms of the distribution of
the momentum jumps $W(\triangle p_k)$ and the Poisson times
$\{t_k\}$.

By performing first the average over $W(\triangle p_k)$ we get
\begin{equation}\label{eq:F01}
F
= \left\langle \prod_{k = 1}^{n}
g(t_{k}) \right\rangle_{\Lambda} ,
\end{equation}
where $g(t_k)=f\left(d \frac{t-t_k}{t}\right)$ and
\begin{equation}\label{eq:g01}
f(x)= \left\langle \exp\left(i x \triangle p\right)
\right\rangle_{\triangle p} = \int du \; W(u) \exp(i x u),
\end{equation}
$\left\langle \cdots \right\rangle_{\Lambda}$ denoting the average
over the shot noise with density $\Lambda$. This is easily computed
\cite{VanKampen}
\begin{eqnarray}
 F &=& 1+\sum_{n=1}^{\infty}\frac{1}{n!}\left[\Lambda\int_0^t
   \left(g(\tau)-1\right) d\tau \right]^n
\nonumber\\
&=&\exp\left(-\Lambda t \int_0^1
   \left[1-f(s d)\right] ds \right) ,
\label{eq:products}
\end{eqnarray}
and yields the visibility (Re denotes the real part)
\begin{eqnarray}
 \cV = \exp (-\Lambda \zeta t), \quad
\zeta = \int_0^1  \left[1-\Re f(s d)\right] ds .
\label{eq:prodvis}
\end{eqnarray}
By using the definition (\ref{eq:g01}), the ``geometrical'' factor
reads $\zeta =
\left\langle1-\sinc(d\triangle p)\right\rangle_{\triangle p}$,
where $\sinc x =\sin x /x$, and
the visibility (\ref{eq:prodvis}) can be given the useful
expression
\begin{equation}\label{eq:prodvis1}
\cV = \exp\left(-\Lambda t \left\langle1-\sinc(d\triangle
p)\right\rangle_{\triangle p}\right).
\end{equation}
Notice that $\cV\le 1$, because $\sinc x \le 1$. Moreover, if the
jumps are symmetrically distributed, i.e.\ $W(\triangle
p)=W(-\triangle p)$, then $f(x)$ is a real function and one can
omit the real part in (\ref{eq:prodvis}), so that $\cV=F$.

\underline{Thermodynamics} -
In order to calculate the visibility from Eq.\ (\ref{eq:prodvis1})
we need to evaluate the kick rate $\Lambda$ and the probability
density of the momentum jumps $W(\Delta p)$ . Planck's blackbody
formula is not valid for small atomic clusters and needs to be
generalized on two counts. One is the finiteness of the number of
modes (freedoms), the other is the reduced stimulated emission.
The former influences the high-energy part of the spectrum, the
latter the low-energy part. In addition, there are finite-size
effects that need to be taken into account. These are usually
dealt with heuristically.

If the interfering cluster can be considered (almost) isolated
during its travel to the screen, its temperature is in general not
in equilibrium with that of the background radiation field. When
photon absorption from the background radiation can be neglected
(which is the case in which we are interested), the photon
emission rate reads
\cite{Hansen2}
\beq
R_T(\omega)=\frac{\omega^2\sigma_{\rm
abs}(\omega)}{\pi^{2}c^{2}}\exp \left[ -\frac{\hbar \omega}{k_B T}
-\frac{k_B}{2C_V}\left( \frac{\hbar \omega}{k_B T}\right)^2
\right],
\label{eq:emission}
\eeq
where $\omega$ is the photon frequency, $\sigma_{\rm abs}$ and
$C_V = Nk_B$ are the absorption cross section and heat capacity of
the small particle, respectively, $T$ its temperature and $N$ the
number of vibrational modes (for example, $N\simeq 170$ for
C$_{60}$, $N\simeq 200$ for C$_{70}$).

The total photon rate reads
\begin{equation}
 \Lambda(T)=\int_{0}^{\infty} d\omega\; R_T(\omega),
\label{eq:J}
\end{equation}
where we assume that the temperature of the molecule does not
change appreciably due to photon emission during the flight. We
will check the validity of this assumption later. In order to
compute these quantities, we have to determine the $\omega$
dependence of the absorption cross section. We shall heuristically
assume the form
\begin{equation}
 \sigma_{\rm abs}(\omega)=a_\ell \omega^\ell ,
\label{eq:sigma}
\end{equation}
where $\ell$ is a positive integer and $a_\ell$ a real number, and
look for the best fit. For instance, in the case of the fullerenes
C$_{60}$ and C$_{70}$, a comparison with experiment \cite{Colin}
yields accurate fits for $a_4= 7.04\times10^{-66}{\rm nm}^2 s^4$
and $a_4 =7.79\times 10^{-66}{\rm nm}^2 s^4$, respectively
\cite{prep}. By plugging Eq.\ (\ref{eq:sigma}) into
(\ref{eq:emission}) one gets the series
\begin{equation}\label{eq:Romegaexp} R_T(\omega)=
\frac{a_\ell\omega^{\ell+2}}{\pi^2 c^2}e^{-\frac{\hbar\omega}{k_B
T}} \sum_{m=0}^\infty \frac{(-1)^m}{(2 N )^m m!}
\left(\frac{\hbar\omega}{k_B T}\right)^{2m}
\end{equation}
and integrating term by term in Eq.\ (\ref{eq:J}) one gets the
asymptotic expansion for large $N$
\begin{equation}\label{eq:Lambdaexp}
\Lambda(T)\sim \frac{a_\ell}{\pi^2 c^2} \left(\frac{k_B
T}{\hbar}\right)^{\ell+3} \sum_{m=0}^\infty \frac{(-1)^m (2 m
+\ell+2)!}{(2 N )^m
 m!}.
\end{equation}
Typical emission rates at $T=2500$ for a time of flight $t \simeq
2$ ms yields 4-5 emitted infrared photons during the flight in the
interferometer. In such case, the temperature of the molecule
decreases by just a few percent, which does not affect the
emission rate, and corroborates our initial assumption [after Eq.\
(\ref{eq:J})].

\begin{figure*}
 \includegraphics{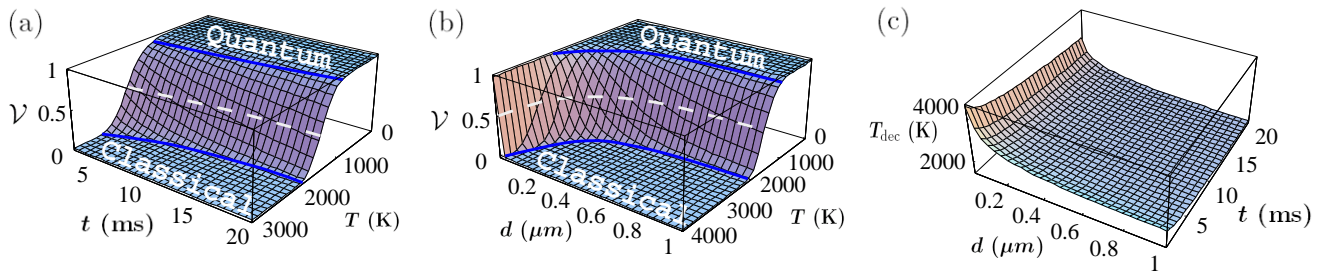}
\caption{(a): Visibility $\cV$ vs $T$ (0--3000 K) and $t$ (3--20
ms) for $d=1\mu$m. (b): Visibility $\cV$ vs $T$ (0--4000 K) and
$d$ (0.01--1 $\mu$m) for $t=10$ ms. The dashed white level line
indicates $\cV=1/2$ and determines the ``decoherence temperature"
plotted in (c): $T_{\rm dec}$ vs. $d$ (0.02--1 $\mu$m) and $t$
(1--20 ms). We plot only the physically relevant ranges of the
parameters.}
\label{Tdec}
\end{figure*}

The momentum kick on the molecule after the emission of a photon
of frequency $\omega$ has magnitude $|\bm p|=\hbar\omega/c$. By
assuming that the emission process is isotropic, the probability
density that the molecule undergoes a momentum jump $\bm p$ reads
\begin{equation}\label{eq:W3D}
W^{(3D)}(\bm{p})=\frac{1}{4\pi \bm p^2} \frac{c}{\hbar \Lambda(T)}
R_T\left(\frac{c |\bm p|}{\hbar}\right),
\end{equation}
from which the one-dimensional probability density can be
evaluated
\begin{eqnarray}
W(\triangle p)&=&\int d^3p\; W^{(3D)}(\bm{p}) \delta(p_x-\triangle
p)
\nonumber\\
&=&\frac{c}{2\hbar \Lambda(T)}\int_{c |\triangle
p|/\hbar}^\infty \frac{d\omega}{\omega} R_T(\omega).
\label{eq:W1D}
\end{eqnarray}
By plugging (\ref{eq:W1D}) into (\ref{eq:g01}) we get
\begin{eqnarray}
 f(x)=\frac{1}{\Lambda(T)}\int_{0}^{\infty}d\omega
  R_T(\omega)
  \sinc\left(\frac{\omega x}{c}\right) ,
\label{eq:gotau}
\end{eqnarray}
and from (\ref{eq:prodvis})
\begin{eqnarray}
\cV(T,d,t )=\exp\left(-[\Lambda(T)-G(T,d)] t \right)
 \label{eq:FoTDt}
\end{eqnarray}
where
\begin{eqnarray}
G(T,d)= \int_0^1 ds \int_{0}^{\infty}d\omega
  R_{\omega}(T)
  \sinc\left(\frac{\omega d}{c}s\right).
\label{eq:GoTd}
\end{eqnarray}
Note that $G(T,d)\sim\Lambda(T)$, for $T\to 0$, so that for low
temperatures $\cV \to 1$. In the high temperature case, on the
contrary, $\cV \to $ 0, as expected for a classical particle.

By plugging the series (\ref{eq:Romegaexp}) into (\ref{eq:GoTd})
and integrating term by term one gets
\begin{eqnarray}
G(T,d)&\sim& \frac{a_\ell }{\pi^2 c d}\left(\frac{k_B
                      T}{\hbar}\right)^{\ell+2}
  \sum_{m=0}^{\infty}\left\{\frac{(-1)^m(2m+\ell+1)!}{\left(2N\right)^m m!} \right.\nonumber \\
  &&\times \left.
  \int_{0}^{\frac{dk_B T}{\hbar c}} dx \frac{\sin \left[ (2m+\ell+2)
  \arctan x \right]}{x(1+x^2)^{(2m+\ell+2)/2}}\right\}. \nonumber\\
  \label{eq:Gexp}
\end{eqnarray}

\underline{Visibility and quantum--classical transition} - By
inserting the expansions (\ref{eq:Lambdaexp}) and (\ref{eq:Gexp})
(with $\ell=4,a_4=7.79\times 10^{-66}$ \cite{prep}) into the
visibility (\ref{eq:FoTDt}) one gets the graphs in Fig.\
\ref{Tdec}(a) for a fixed distance between the slits, and in Fig.\
\ref{Tdec}(b) for a fixed time of flight. A quantum system,
characterized by the value $\cV=1$, tends to display a classical
behavior, characterized by $\cV=0$, when the time of flight and/or
the distance between the slits are increased. This
quantum-classical transition takes place at a ``decoherence
temperature" $T_{\rm dec}$ determined by the level curve
\begin{equation}
\cV = 1/2 \quad \Longleftrightarrow \quad [\Lambda(T)-G(T,d)] t- \ln 2=0.
\end{equation}
$T_{\rm dec}(d,t)$ is plotted in Fig.\ {\ref{Tdec}}(c). The
transition between the quantum and classical behavior is very
sharp, both in Figs.\ {\ref{Tdec}}(a) and {\ref{Tdec}}(b), and
this enables us to define the decoherence temperature in a
clear-cut way. These graphs are the central results of our
analysis.

A mesoscopic system, such as a macromolecule, can be attributed a
temperature, in the sense of Eq.\ (\ref{eq:emission}), by virtue
of its large number of freedoms $N$. In a double slit interference
experiment, the degree of freedom associated with the interfering
pattern [the relevant variable being $x$, see Eq.\
(\ref{eq:initial})] plays a special role. We now argue that, in
general, such an ``interfering" freedom is not in the same thermal
state as the others.

Let the experiment last for a time $t$ (the time of flight of the
molecule in the interferometer) and $H_{\rm exch}$ be the
interaction Hamiltonian responsible for the coupling between the
interfering freedom and the environment. Interference can be
observed if
\begin{equation}
\left\langle \int_0^{t} H_{\rm exch} dt \right\rangle_T \lesssim
\hbar \ll kT t,
\label{eq:actions}
\end{equation}
where the average $\langle \cdots \rangle_T$ is taken over the
initial state of the total system
\footnote{We are always neglecting the temperature of the photon
bath, see comments after Eq.\ (\ref{eq:initial}): strictly
speaking our environment is made up of two parts (the other
freedoms of the molecule + the photon bath), that are not in
equilibrium, and $T$ is the ``local" temperature of the
rotovibrational and electronic modes of the macromolecule,
responsible for photon emission. We are not considering the (much)
slower equilibration process of the whole environment.}. The above
one is a condition on the exchanged action and the environmental
temperature. For example, if the average in Eq.~(\ref{eq:actions})
is understood in the r.m.s.\ sense and
$H_{\mathrm{exch}}=-\xi(t)x$, like in (\ref{eq:Hamiltonian1}), we
have $\left\langle \int H_{\rm exch} dt \right\rangle
=\sqrt{\Lambda t} \langle \triangle p \rangle  d = \triangle
p_{\rm tot} d \le \hbar$ where $\triangle p_{\rm tot}(T)$ is the
total recoil due to a momentum random walk
\cite{fullerene}. In such a case the first inequality in
(\ref{eq:actions}) is nothing but Heisenberg's inequality and this
clarifies the rationale behind it. When this condition is
satisfied, the macromolecule interferes. During the interference
experiment, energy flows between the environment and the
interfering ``colder" freedom. Such a freedom, associated with the
interfering component $p_{x}$, approaches equilibrium (at a
temperature $T$), via momentum---and energy---transfer, during the
momentum random walk process described above: eventually,
visibility vanishes and interference is lost when the first
inequality in (\ref{eq:actions}) ceases to be valid. This is a
rather fast process, that induces a classical behavior in the
(relevant interfering variable of the) mesoscopic system. The
thermalization process sets in afterwards, when $\left\langle
\int H_{\rm exch} dt \right\rangle \simeq kT t \gg \hbar$
instead of (\ref{eq:actions}), and is much slower. This is,
altogether, a remarkable picture, that adds spell to the
interfering features of these mesoscopic systems, as well to the
many additional problems that must be considered
\cite{Hegerfeldt} in order to get a complete picture of these
phenomena.

\textbf{Acknowledgements.} We thank M.\ Arndt, S.\ Kurihara, I.\
Ohba and A.\ Zeilinger for interesting remarks.


\end{document}